\documentclass[12pt]{article}
\usepackage{epsf}
\begin{document}

\title{A housing-demographic multi-layered nonlinear model to test regulation
strategies}

\author{Ram\'{o}n Huerta$^{a}$, Fernando Corbacho$^{b,c}$, 
Luis F. Lago-Fern\'{a}ndez$^{b,c,}$\footnote{Corresponding author. E-mails: rhuerta@ucsd.edu (R. Huerta), fernando.corbacho@cognodata.com (F. Corbacho), luis.lago@uam.es (Luis F. Lago-Fern\'{a}ndez).}\\\\
$^{a}$ Institute for Nonlinear Science, University of California, San Diego,\\ 
La Jolla, CA  92093-0402\\
$^{b}$ Escuela Polit\'{e}cnica Superior,
Universidad Aut\'{o}noma de Madrid,\\ 28049 Madrid (SPAIN)\\
$^{c}$ Cognodata Consulting, Caracas 23, 28010 Madrid (SPAIN)\\
}


\maketitle

\begin{abstract}

We propose a novel multi-layered nonlinear model that is able to capture and predict 
the housing-demographic dynamics of the real-state market by simulating
the transitions of owners among price-based house layers. This
model allows us to determine which parameters are most effective to
smoothen the severity of a potential market crisis.
The International Monetary Fund (IMF) has
issued severe warnings about the current real-state
bubble in the United States, the United Kingdom, Ireland, the Netherlands,
Australia and Spain in the last years. 
Madrid (Spain), in particular, is an extreme case of this
bubble. It is, therefore, an excellent test case to analyze housing
dynamics in the context of the empirical data provided by the Spanish {\em
National Institute of Statistics} and other
sources of data. The model is able to predict the mean house occupancy, 
and shows that i) the house market conditions in Madrid are
unstable but not critical; and ii) the regulation of
the construction rate is more effective than interest rate changes. Our
results indicate that to accommodate the construction rate to the total
population of first-time buyers is the most effective way to maintain
the system near equilibrium conditions. In addition, we show that to
raise interest rates will heavily affect the poorest housing bands of the
population while the middle class layers remain nearly unaffected.
\end{abstract}

\section{Introduction}

 A crisis on the housing market
could heavily reverse the current positive economical indicators \cite{crashbook,garber2000} and
have a negative general impact in the world economy, as shown in the
Asian crisis in 1997 \cite{asian}.  Real-state prices have increased
more than fifty percent in Australia, Ireland, United Kingdom and
Spain from 1997 to 2004. These increases are hardly explainable in terms of economic
fundamentals alone, even including record-low interest
rates \cite{imf2}. Spain, in particular, holds high risk figures in
terms of ratios of house prices to disposable income per worker (RPI)
and house prices to rent (RPR). According to IMF calculations Spain
has $3.6$, $3.6$, $2.5$, $2.2$, $1.8$ times the RPI of Germany,
Japan, United States, France and United Kingdom, respectively
\cite{imf2}. Despite the overwhelming cost to acquire a house
in Spain compared to all developed countries, house unit prices 
currently remain unaffected. In addition, in terms of the RPR, which is a good
estimation of the yield of the investment, Spain has $3.42$, $3.32$,
$1.9$, $1.8$, and  $1.28$ times the RPR of Germany, Japan, France, United
States and United Kingdom, respectively \cite{imf2}.
The RPI index has been proved to be a good
reference of the distance to market equilibrium
\cite{overprize}. Therefore, Spain is a valid
scenario to analyze the conditions to understand and control a housing
market by means of nonlinear dynamical models.

In this paper we develop a mean field model of the house occupancy derived from 
a stochastic process as previously used in the study of epidemic dynamics \cite{epidemic}. This nonlinear model is employed as a tool to ascertain the possibility  
of sudden changes in the dynamics as a function of the control parameters.
The model is based on dissecting the house population in several groups or layers.
Each layer has a given number of houses, which can be 
either occupied or empty. Every family occupying a house in a given layer
has a certain probability rate to migrate to any other layer.
In addition, there exists a base pool of nonowners, which can jump to the housing
layers at a certain probability rate. Thus the model
captures the dynamics of the mean migrations between housing layers.

As a particular case, we analyze the house market in the city of Madrid
in the framework of the model. The parameters of the model have been obtained 
from different sources such as the Spanish National Institute of 
Statistics, the Spanish Ministry, and 
some valuation companies \cite{censo91,epa,ine,tasacion}. 

Our main results may be summarized as follows: 

\begin{itemize}

\item The model is able to predict the occupancy levels in 2001 given
  the parameters obtained from 1991 data. This is a good indication
  that the mean description is able to model the real dynamics. 

\item The model shows that the house market for the city of 
Madrid asymptotically evolves to an out-of-equilibrium
condition. It is rather worrisome that the new housing units cannot 
be replenished by first-time buyers.

\item Critical phenomena do not exist in the context of this model. Only
smooth changes can be expected.

\item A sudden increase in the interest rate will seriously affect the 
occupancy level of the lowest layer of the market, i.e. the poorest 
housing unit sector, while the middle layers remain nearly unaffected. 

\item According to the model, the most effective way to control 
the out-of-equilibrium situation of the Madrid house market 
is to down regulate the construction rate. 

\end{itemize}

There is a large body of research in the housing market
\cite{review,review2}, most of the analyses aiming at predicting
house prices. We do not intend to estimate prices, 
our modeling efforts fit better with housing-demographic
models \cite{population1,population2}. The novelty of our approach 
lies on building a multi-layered nonlinear dynamical model that includes family
migrations across layers of housing units. The whole population
is separated in non-overlapping housing bands, which are
estimated from census data. Our model neglects 
the random component of the stochastic process, because we use 
a mean field approach under the assumption of random mixing. The
random mixing approximation basically states that any family 
can, in principle, uniformly access any house. 
This random mixing approximation allows to obtain a set of ordinary
differential equations (ODEs) derived from a stochastic process.  
The formalism that we use here sets us apart from previous approaches.

\section{Overview of the model}

In this section we provide an overall description of the non-linear dynamical model for 
the house occupancy. The model equations will be explicitly derived in the next section. We divide the total housing population into different price bands or layers, and model the transitions of owners among these layers. Note that, in this paper, we only use the price as a tool to determine the different house bands. Let us assume that there are $N_{layers}$ layers. We define $N_{i}(t)$ as the {\it total} number of houses, or units, in layer $i$ at time $t$. Each of the layers has a certain 
number of {\it occupied} houses, $O_{i}(t)$. These are defined as houses occupied by owners. In addition to the house price bands, we consider a source of new buyers, which we call the {\it base pool}. It accounts for non-emancipated people, immigrants, and rented units. 

A group of people living in the same house  will be called a 
{\it family}. The family is the basic people unit in our model, and so we are modeling 
family jumps among house layers. Note that the number of families in a given layer $i$ 
equals the number of occupied houses in that layer, $O_{i}(t)$. 
The concept of family applies to non-emancipated people 
and immigrants as well. However, in these cases information concerning the total number of 
single individuals is more frequently available in census databases, so we must apply a 
correction factor when calculating the number of families ($\Sigma(t)$) in the base pool.

\begin{figure}[ht!]

\parbox{0.42\textwidth}{
\centerline{
\epsfxsize=0.42 \textwidth
\epsfbox{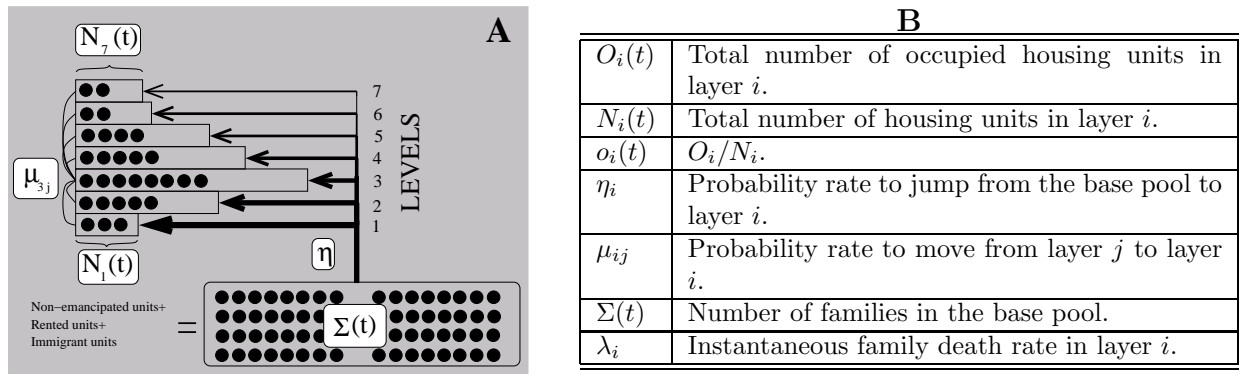}
}
}
\parbox{0.6\textwidth}{
\begin{center}
\footnotesize
{\normalsize \bf B}\\
\begin{tabular}[c]{|l|p{0.43\textwidth}|}
 \hline
\hline
$O_i(t)$ & Total number of occupied housing units in layer $i$.\\
\hline
$N_i(t)$ & Total number of housing units in layer $i$.\\
\hline
$o_i(t)$ & $O_i/N_i$.\\
\hline
$\eta_i$ & Probability rate to jump from the base pool to layer $i$.\\\hline
$\mu_{ij}$ & Probability rate to move from layer $j$ to layer $i$.\\\hline
$\Sigma(t) $ & Number of families in the base pool.\\\hline
$\lambda_i$ & Instantaneous family death rate in layer $i$.\\ \hline \hline
\end{tabular}
\end{center}
}							
\caption{\footnotesize (A) Explanatory figure of the model and the main parameters required 
in our  housing-demographic model. Each layer represents a price housing band. The dark
circles represent occupied housing units. $\Sigma (t)$ represents the pool of units that can
enter any of the vacant sites on any of the layers with some probability rate ${\bf \eta}$.
Each of the layers have a total number of housing units $N_{i}(t)$. The transition probability
rate $\mu_{3\, j}$ denotes the probability of having transition from layer $j$ 
into layer $3$. Note that we omitted the rest of the transitions for the sake of the 
readability of the figure. (B) Glossary of variables and parameters of the nonlinear
  dynamical model.
\label{fig2}}
\end{figure}

The migrations among house layers are modeled in terms of transition probabilities. We
denote by $\mu_{ij}$ the probability rate for a family in house layer $j$ to move to
house layer $i$. Equivalently, we denote the probability rate to move from the base pool 
to any of the house layers $i$ by $\eta_{i}$. Finally, there is a
probability for any family to disappear, or die, leaving its house empty. We call this probability
the {\it death rate}, $\lambda_{i}$, which we assume to be constant for each layer $i$.
Figure \ref{fig2}A shows an overall scheme of the model; figure \ref{fig2}B displays a list of the variables and parameters involved.

With all the above ingredients we pose a set of non-linear coupled equations for the mean
occupancy of each level: $o_{i}(t) = O_{i}(t) / N_{i}(t)$. The model equations are derived from
the mean field approximation to a stochastic process, as used in epidemic dynamics. 
Details are provided in the next section.

\section{Derivation of the model equations}
The model proposed here is similar in derivation to the ones used in epidemic modeling \cite{epidemic,huerta} where 
the stochastic epidemic process is reduced to a set of ordinary differential equations (ODEs).
These ODEs capture quite faithfully the behavior of the stochastic process behind epidemics. The main advantage 
of this approach is that the complexity of the process is simplified to a formalism that allows an easier understanding of the qualitative behavior. The parameters can be easily related to the 
end result of the stochastic process. In most cases the ODEs match well the stochastic process, although there
are some others where the ODE description fails, for example, for finite-size effects. Overall, the ODE description
is a very good framework to gain understanding that can complement very well stochastic modeling. Our 
main contribution is to bring these convenient tools to housing-demographic modeling. 

To model the dynamics across housing layers we use two possible states for any house: 
occupied and empty. The total number of houses in each layer $i$,
$N_{i}(t)$, evolves in time according to a function estimated from the census data. 
Given the number of occupied units in layer $i$, $O_{i}(t)$, and the transition
probability rate from layer $j$ to layer $i$, $\mu_{ij}$, the probability that 
a family jumps from layer $j$ to layer $i$ in the time interval $dt$ is 
$\mu_{ij}\left\{1-O_i(t)/N_i(t)\right\}\, dt$. This probability already assumes that 
a family can only occupy an empty house in layer $i$. This is equivalent to the random 
mixing approximation widely used in epidemic modeling \cite{epidemic}. The net flow
into layer $i$ is the difference between the number of incoming and outgoing families
in the time interval $dt$:

$$
F_{i}=\sum_{j=1}^{N_{layers}} \left[O_j(t) \mu_{ij}\left\{1-\frac{O_i(t)}{N_i(t)}\right\} -O_i(t) \mu_{ji}\left\{1-\frac{O_j(t)}{N_j(t)}\right\}\right] dt
$$

This flow equals the variation in the number of occupied states in layer $i$ during 
the time interval $dt$, so we can write the following set of ordinary differential
equations (ODEs) for the layer occupancy:

$$
\frac{d O_i (t)}{dt}=\sum_{j=1}^{N_{layers}} \left[\mu_{ij} O_j (t)\left\{1-
\frac{O_i (t)}{N_i(t)}\right\} -\mu_{ji}O_i (t)\left\{1-\frac{O_j (t)}{N_j(t)}\right\}\right]
$$

It is critical to include the dynamical contribution of the base pool. The 
probability for a family to jump from the base pool to the layer $i$ in the time 
interval $dt$ is given by $\eta_i \left\{1-(O_i(t)/N_i(t))\right\} \, dt$. Then, the net 
change in layer $i$ due to the flow from the base pool is simply
$ \Sigma (t) \eta_i \left\{1-(O_i(t)/N_i(t))\right\} \, dt$. Finally, we will consider a 
family death rate for each layer, $\lambda_i$, which contributes to the variation
in layer occupancy with the term $-\lambda_i O_i(t) dt$. The final model equations
can be written as:  

\begin{equation}
\frac{d O_i}{dt}=\eta_i \left(1-\frac{O_i}{N_i(t)}\right)\Sigma(t)- \lambda_i O_i+
\sum_{j=1}^{N_{layers}} \left[\mu_{ij} O_j\left\{1-\frac{O_i}{N_i(t)}\right\} -\mu_{ji}O_i\left\{1-\frac{O_j}{N_j(t)}\right\}\right]
\label{basic}
\end{equation}

\noindent Since we plan to analyze the asymptotic behavior of the equations, 
we define a new variable, $o_i(t)\equiv O_i(t)/N_i(t)$, which is the normalized occupancy level, 
bounded between $0$ and $1$. We can rewrite the set of equations \ref{basic} in terms of
the normalized occupancy as:

\begin{equation}
\frac{d o_i}{dt}=\eta_i (1-o_i)\frac{\Sigma(t) }{N_i(t)}-o_i\frac{d}{dt}\log N_i(t)-\lambda_i o_i+\sum_{j=1}^{N_{layers}} \left(\mu_{ij} o_j (1-o_i)\frac{N_j(t)}{N_i(t)} -\mu_{ji}o_i (1-o_j)\right)
\label{normalized}
\end{equation}

There are three terms in these ODEs with explicit dependence on time. The first one is the drive from the base pool
to saturation levels. If $\Sigma (t)$ grows faster than single layers do, then all layers will saturate. 
The third term with explicit dependence on $t$ is multiplied by $N_j(t)/N_i(t)$. This term implies that, if the size of layer $j$ grows much faster than layer $i$, in the asymptotic limit the layer $i$ will be totally full, with a huge demand in that layer that will quickly change the band location in the whole distribution of layers.

\section{Test case: the city of Madrid}

Madrid is a particularly extreme case of the real-state bubble
in Spain \cite{imf2}. This fact, together with the availability of data to estimate model parameters, 
makes Madrid an interesting test case to be analyzed in the context of our mean field model. 
The two main sources of data used to feed the model parameters are the Spanish National 
Institute of Statistics (INE) and the Spanish Ministry, but we have also used 
data provided by valuation companies and real-state internet sites. First we will provide an overall description of the model parameter estimation in section \ref{sec_parameters}. Then, in section \ref{sec_full_model}, we will use these parameters in the model equations (\ref{normalized}) to understand the implications of current market conditions.

\subsection{Parameter estimation}

\label{sec_parameters}

\begin{figure}[ht!]

\centerline{
\epsfxsize=0.9 \textwidth
\epsfbox{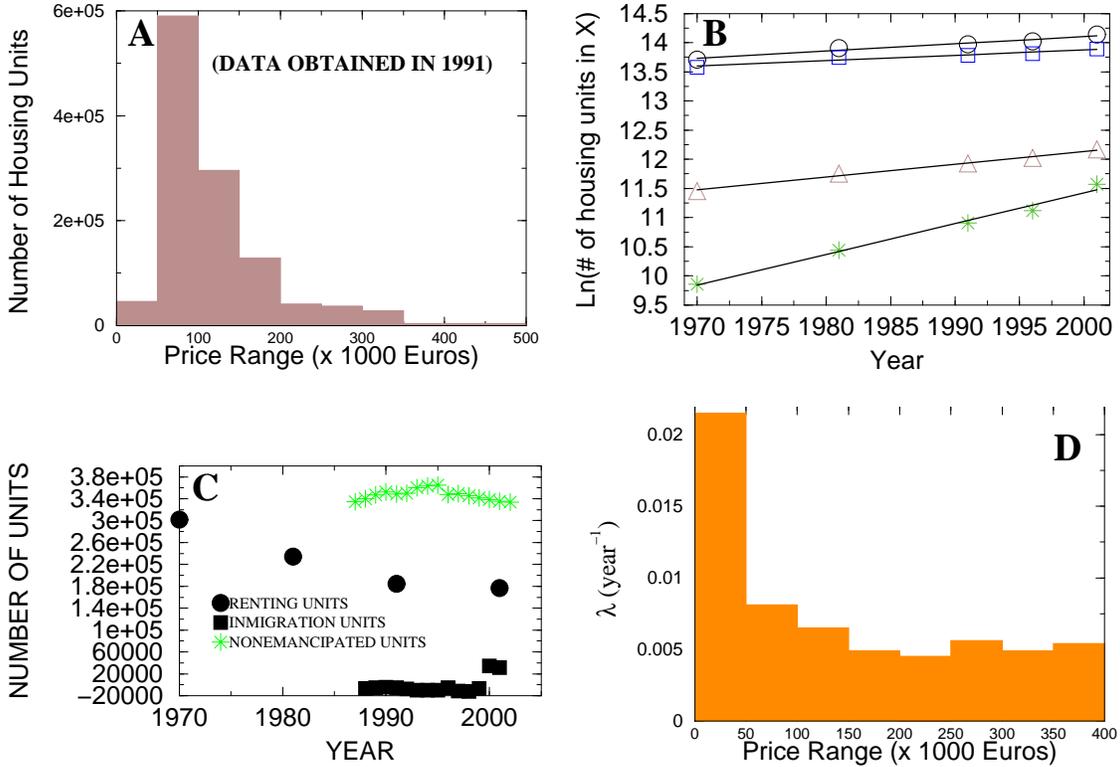}
}

\caption{\footnotesize (A) Distribution of house prices in the city of Madrid (Spain) 
during 1991, extracted from data provided by \cite{censo91} and \cite{tasacion}). 
(B) Fits to the logarithm of the total number (circles), primary (squares), 
secondary (triangles), and empty (stars) houses. The data is provided for only 5 points in time 
\cite{ine}. The fits yield the following slopes: $0.012\pm0.001$ year$^{-1}$ (total), 
$0.009 \pm 0.001$ year$^{-1}$(primary), $0.022 \pm 0.001$ year$^{-1}$(secondary), and 
$0.05 \pm 0.003$ year$^{-1}$ (empty). In all the cases the correlation factor of the 
linear regression is greater than  $0.98$. (C) Time evolution of the different subsets 
of the base pool: non-emancipated families (stars), immigrant families (squares), 
and renting families (circles). (D) Instantaneous death rate per layer, $\lambda_i$. 
Parameters estimated from 1991 data.}
\label{fig1}
\end{figure}

First of all we must determine the price layers. Figure \ref{fig1}A represents 
the distribution of house prices in the city of Madrid in 1991 (data from
\cite{censo91,tasacion}). 
This distribution provides the total number of houses per 
layer, $N_{i}$, using $7$ layers that account for the $99\%$ of the total number
of houses. The number of occupied houses in each layer, $O_{i}$, is also 
provided by \cite{censo91,tasacion}. To choose the price size
of each layer we find a compromise between two opposing criteria: i) the maximum
number of layers in order to have a detailed distribution of the housing stock; and ii) 
the widest price size per band such that the transition probability rates between layers are not very small when normalized to the integration time scale. This second criterion intends to avoid finite size effect problems. 

The time evolution of the number of houses per layer, $N_i(t)$, is hardly available in public 
databases. Nevertheless, as shown in fig. \ref{fig1}B, the total population of 
houses is available at five different years since 1970 \cite{ine}. In that figure we can
see that the total population of houses is very well fit by an exponential function of 
time. 
We will assume that the shape of the layers distribution is time invariant, and will 
apply the same exponential growth to all layers. This assumption is supported by two facts: 
(i) the evidence of self-regulation of each of the 
layers as shown in \cite{regulation} for the city of Philadelphia, {\em i.e.},
undervalued houses compared to similar type of houses get more appreciated than the average; and (ii)
the similarity of price distributions in the cities of Pitt County 
(North Carolina) \cite{pitt} and Madrid (figure \ref{fig1}A).  

The number of families in the base pool ($\Sigma (t)$), {\em i.e.}, the subset of families that do not own a property, is estimated from the INE databases 
as the sum of: (i) the number of people that are not emancipated within the range
of age where people usually emancipate (data from \cite{epa,objovi}); (ii) the number 
of families that rent an apartment (data provided by \cite{ine}); and (iii) the number of 
incoming families due to the immigration (data from \cite{ine}). These statistics do 
not provide family units, but they report single people data instead. So a headship 
factor of $2$ has been applied when calculating $\Sigma (t)$.
In fig. \ref{fig1}C we can see the time evolution of the three subsets that conform 
the base pool. The two main contributions to the pool have been decreasing 
in the last few years. On the other hand the immigrant population has undergone a sharp 
increase. However, in contrast to the popular view, this subset of the population is 
not officially large, and it might just compensate for the negative tendency in the base pool size 
time evolution. In contrast to the total number of houses (fig. \ref{fig1}B), there is 
no exponential growth of $\Sigma (t)$. In fact, we will assume it is a constant. This 
situation, if maintained, would lead the system to complete depletion as we will discuss 
bellow in the context of the model. 

Other important parameters to be estimated are the transition probability rates from
the base pool to the different layers, $\eta_{i}$, and the transition probability rates 
among layers, $\mu_{ij}$.
The average transition probability rate from the base pool to any of the layers, $\bar{\eta}$, can be estimated from the total number of new occupancies in 1981, 1991 and 2001, as follows. 
The total number of occupied houses in 1981 was $699,557$, in 1991 was $789,444$ and 
in 2001 was $908,790$ \cite{ine}. Therefore the occupancy rate from 1981 to 1991 was 
$8,989$ new occupancies per year, and from 1981 to 1991 it was $11,935$ new occupancies per year.
The base pool population in 1981 is estimated in $568,300$ families (we have 
data for the non-emancipated population since 1986 and we apply a headship per future household of $2$),
in 1991 it was $532,717$ and in 2001 it was $543,323$. Then an estimation of the probability rate 
of having one base pool family accessing any of the layers is $0.016$ year$^{-1}$ from 1981 to 1991, 
and $0.022$ year$^{-1}$ from 1991 to 2001. This rate has been increasing 
in the last few years, maybe due to the decrease in the interest rates.

\begin{figure}[htb!]

\centerline{
\epsfxsize=12cm
\epsfbox{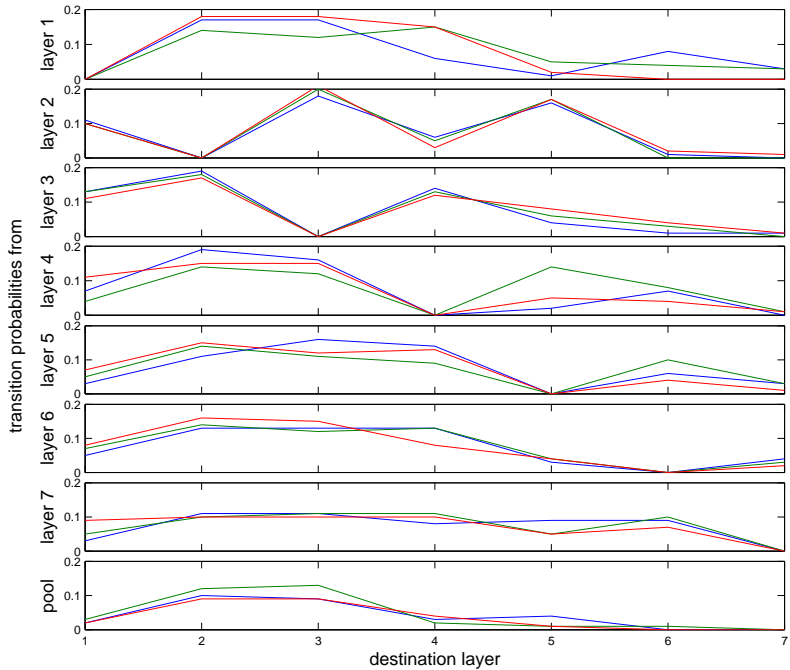}
}						
\caption{\footnotesize Solutions of three different runs of the genetic algorithm used to estimate 
the transition
probabilities. These plots show the values of the transition probabilities from any of
the layers including the base pool (rows, $y$ axis) to any of the $7$ property layers
($x$ axis) in ten years. To get the probabilities in one year the values must be divided
by $10$. The standard deviation with respect to the target value $\hat{{\bf o}}$ is 
in all the cases less than $10^{-4}$.}
\label{fig_probs}
\end{figure}

To fit the probabilities $\mu_{ij}$, we assume that the occupancy levels in $1991$ are 
stationary, and search for the values of $\mu_{ij}$ that provide these levels 
at equilibrium. The available data do not permit direct calculation of these 
probabilities, nevertheless it is possible to estimate their average by extrapolating 
the number of housing transactions in Spain (provided by analysis office by 
\cite{bbva} and \cite{objovi}) to the city of Madrid. 
If we take this total number of transactions and discount the probability of transition 
from the base pool and the estimation of housing transactions
due to investments, we obtain $\bar{\mu}=0.052$ year$^{-1}$. This basically states that the 
time it takes a family to change to a new house is 20 years on average for the city of Madrid. 
This is imposed as a constraint in the following calculations.
 Therefore, we will assume that the occupancy levels are 
slightly off from the equilibrium state and that the growth of the population is compensated 
with the growth of the total number of housing units. The occupancy level of each of the $7$
layers at 1991 is given by $\hat{{\bf o}}= (0.77, 0.56,  0.68,  0.62,  0.85, 0.62, 0.39)$. 
The parameter search is performed by means of a genetic algorithm with a fitness function 
that measures the Euclidean distance between $\hat{{\bf o}}$ and the model equilibrium 
solution ${\bf o}$. To reduce the search space we apply two constraints: (i) the average 
value of $\mu_{ij}$ is, as stated above, $\bar{\mu}=0.052$ year$^{-1}$; and (ii) the 
derivative between close parameters is as small as possible. A total of $50$ different 
simulations were run, the outcomes of three of them are shown in figure \ref{fig_probs}.
Although the specific values of the probabilities do not match for the three simulations, the shape and tendency maintain qualitative agreement. We use their average value when solving the model equations.

The last important estimation in our model is the family death rate for each layer. We have data
of the absolute number of deaths per age and the age distributions per housing layer \cite{ine}.
With these data, assuming a family is composed of two people very close in age, we 
can calculate the instantaneous family death rate per layer (fig. \ref{fig1}D). 
We can see that the lower layers have a higher death rate, because the 
average age of families in these layers is higher. This can be explained by the slow integration of the old houses into the lower price layers. On the other hand, newly constructed houses tend to fill higher layers.

\subsection{Results}

\label{sec_full_model} 

\begin{figure}[ht!]

\centerline{
\epsfxsize=0.9\textwidth
\epsfbox{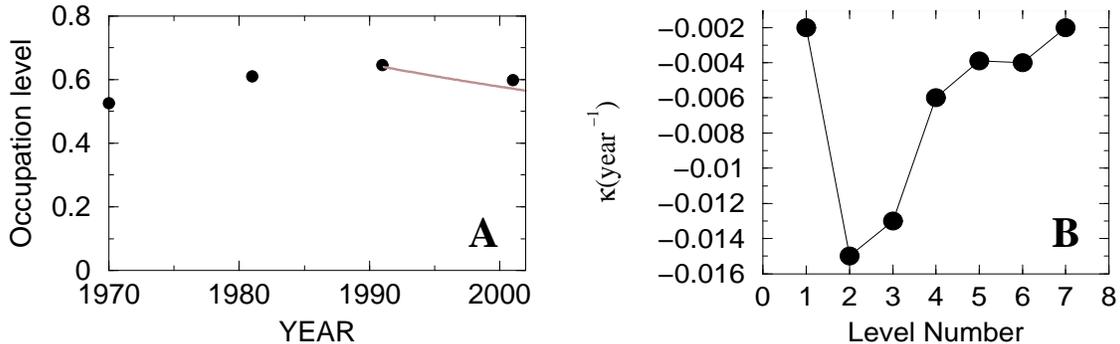}
}

\caption{\footnotesize A. Occupation level for 4 different years. The occupation level
has been obtained as the sum of paid property, property with mortgage, and property passed 
from parents to sons and daughters divided by the total number of housing units. The solid 
line represents the prediction of the model by using 1991 data. B. Decay rate of the occupation
for each of the layers. This decay rate has been obtained by fitting the $o_i(t)\propto exp(\kappa_i \, t)$
from 1991 top 2010. 
\label{fig5}}
\end{figure}

For the problem under analysis, and taking into account the exponential growth of the total number of houses for the city of Madrid, the model equations (\ref{normalized}) can be rewritten as:

\begin{equation}
\frac{d o_i}{dt}=\eta_i (1-o_i)\frac{\Sigma }{\hat{N}_i \exp{k\, t}}-o_i \, (k +\lambda_i)+\sum_{j=1}^{N_{layers}} 
\left(\mu_{ij} o_j (1-o_i)\frac{\hat{N}_j}{\hat{N}_i} -\mu_{ji}o_i (1-o_j)\right)
\label{eq_madrid}
\end{equation}

\noindent where $k$ is the constant of the exponential growth fitted in figure \ref{fig1}B, and $\hat{N}_i$ is the total number of houses in layer $i$ in 1991 (figure \ref{fig1}A). Integrating the equations with the parameters of previous section, we can make predictions on the mean occupancy level. In particular, the decreasing trend of the occupancy levels from 1991 to 2001 is well predicted by the model using data from 1971 to 1991 (see figure \ref{fig5}A).

A sufficient condition to keep the system asymptotically away from $0$ (empty) or $1$ (full) is the following

\begin{equation}
N(t)=\left(N(0)+\int_{0}^{t}\left(\sum_i {\eta_i}\right) \frac{1-\bar{o}}{\bar{o}}\,e^{\bar{\lambda}\, \tau} 
\Sigma (\tau)\, d\tau\right)e^{-\bar{\lambda}\,t},\label{eqsimple}
\end{equation}

\noindent where $\bar{o}$ is the mean occupancy level, and $\bar{\lambda}$ is the mean instantaneous household death rate. If the time evolution 
of the base population is constant, as it appears according to figure \ref{fig1}C, then  $N(t)\propto exp(-\bar{\lambda} t)$.
The optimal equilibrium occupancy levels in the housing market is out of the scope of this paper. Nevertheless it is certainly desirable 
to steer the system to a equilibrium condition whose asymptotic state is far from the $1$-$0$ extremes. 
Equation (\ref{eqsimple}) intends to 
provide a simple framework for policy makers to regulate the system in a smooth way from a housing-demographic 
perspective, disregarding price value of the housing units.

Nevertheless, although equation (\ref{eqsimple}) is useful due to its simplicity, the intrinsic dynamics of the system is more complicated. Each layer follows a different trend due to the integration of the ODEs. In order to provide a vacancy rate for each of the layers, {\em i.e.}, the pace at which each layer becomes empty, we fit an exponential function of time to the predicted occupancy levels, such as
$o_i(t)\propto exp(\kappa_i \, t)$, for 20 years since 1991. Note that $o_i(t)$ is not an exponential decay function, but since we are mostly interested in short periods of time ($10$ to $50$ years), the exponential fits are adequate. As it can be seen in Fig. \ref{fig5}B, layers 2 and 3 undergo the heaviest vacancy rate at current market conditions, while the other layers are in a more stable situation. This indicates that the middle class layers are subject to the heaviest speculative process, while the other layers may have a lower demand.

\section{Economical conditions and temperature control parameter}

The economical conditions that may affect the transition probabilities among layers are incorporated into the model by means of a single parameter $T$, that plays a role similar to the temperature in statistical physics. For example, if the interest rates are raised, the probability of transition from one layer to the rest or from the base pool to any layer is decreased, which is modeled by a temperature decrement.
Another example, if the RPI grows excessively, then the temperature is also lowered, with the consequent decrease of the transition probabilities. Unfortunately, we do not have data to quantify the dependence of transition probabilities on economical
conditions. To be able to estimate them we would need data that indicates the number of mortgage requests and the amount loaned as a function of time. This has to be provided 
by banks and we currently do not have such data. The general dependence of the transition probability rates are therefore unknown. Initially, we assume that the transition rates ($\mu_{ij}$ and $\eta_i$) have a linear dependence on the parameter $T$. Therefore, the parameters $\mu_{ij}$ and $\eta_i$ in equation \ref{eq_madrid} are replaced by $T \mu_{ij}$ and $T \eta_i$. This dependence basically implies that the economical conditions uniformly affect all the probability rates in the same way. The goal is to determine whether there is a value of the temperature parameter that leads to a sudden change on the occupancy levels. As long as the dependence of the probability rates on the temperature is an analytic function, the existence of critical behavior in the dynamical system  should be preserved. In other words, if $\mu_{ij}(T)$ and $\eta_i(T)$ are discontinuous, then the criticality emerges from parameter dependence on $T$, not from the dynamical system under consideration.
In this work, we can only concentrate on the dynamical system behavior.

\begin{figure}[ht!]

\centerline{
\epsfxsize=0.9\textwidth
\epsfbox{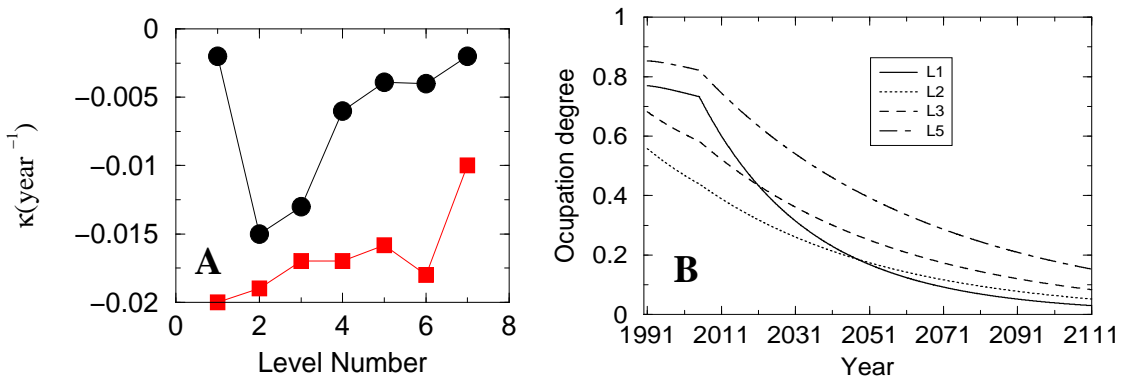}
}

\caption{\footnotesize (A) Decay rate of the occupancy level
for each of the layers. This decay rate has been obtained by fitting the $o_i(t)\propto exp(\kappa_i \, t)$. Circles represent current market conditions. Squares represent a hypothetical situation where the probability rates are reduced 10\%. (B) Effect of a sudden decrease in the temperature in 2005 using the parameter conditions 
estimated for 1991. Layer 1 is the most sensitive to changing economical conditions, while layers 2 and 3 are nearly unaffected. Note that, for clarity reasons, we have not included higher layers because of their similarity to layer 5.
\label{temperature}}
\end{figure}

First, to determine
the effects of freezing the system by temperature reduction we fit the decay rate of the occupancy level to a exponential function as shown in the previous section (see Fig. \ref{temperature}A). 
The middle layers are more robust to temperature changes. The middle layers show more resilience to critical economical
conditions while both extremes of the multilayer model are highly sensitive. This contrasts with the depletion resistance observed in Fig. \ref{fig5}B 
for layer one. Thus, the lowest layer is the most sensitive to changes in the
economical conditions.  

As an example of the sensitivity of the lowest layer we introduce in 2005 a sudden decrease in the temperature
that can be seen as a strong increase in the interest rates. In Fig. \ref{temperature}B we can see
that a sudden change in the temperature yields a sharp slope modification in layer 1. On the 
other hand, it is striking to 
see that layers 2 and 3 are not as heavily impaired. 

In summary, using the 1991 market parameters and the nonlinear dynamics formalism 
proposed in this article, the housing market of Madrid is not subject to critical (non-continuous) behavior on the temperature control parameter. It is clear that 
layers (2 and 3) are very resilient to temperature changes. Nevertheless, although criticality is absent, layer 1 undergoes the most rapid changes to variable economical conditions. The layers with most expensive houses can potentially undergo serious reversals under difficult economical conditions. By means of this multilayer model we can identify which sectors of the housing population would be more affected. An overall
regulation of the market following equation (\ref{eqsimple}) may not lead to uniform stabilization
of the system. Multi-layer models can be used to detect the individual impact of global policies.

\section{Testing regulation strategies}

\begin{figure}[ht!]
\centerline{
\epsfxsize=4in
\epsfbox{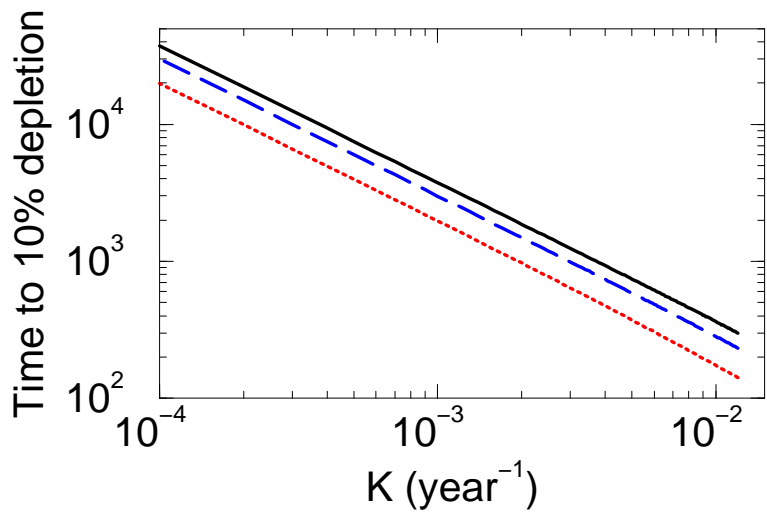}
}						
\caption{\footnotesize Time to 10\% depletion as a function of a sudden decrease of the housing unit growth factor in 2005 
using the parameter conditions 
estimated for 1991. The solid line corresponds to layer 1, the dotted line to layer 2, and the dashed line to layer 4. The rest 
of the layers follow the same power law.
\label{Kfactor}}
\end{figure}

The temperature parameter is not a strong control parameter, since it does not lead the system into an asymptotic equilibrium. Therefore, the question of whether there is a parameter that can globally keep the system in a nearly asymptotic equilibrium remains unanswered. A possible candidate is the growth rate of the number of houses, $k$.
As we have seen in section 4 and fig. \ref{fig1}B the house growth rate is exponential 
for the city of Madrid, which is in contrast with the nearly flat increase of the base pool population (fig. \ref{fig1}C). This situation is not sustainable, and the model can be useful to determine what to expect when regulating the  house growth. We have integrated equations \ref{eq_madrid}
with the parameter values obtained for 1991, suddenly changing the growth factor $k$ at 2005.
Time to $10\%$ depletion versus $k$ is shown in figure \ref{Kfactor}.
As we can see, there is a power law dependence. Basically, the time 
required to a 10\% depletion is proportional to $(1/k)$ (fits actually yield 
$k^{-1.01}$ to $k^{-1.04}$). This simple solution allows to easily estimate what the 
effects of house growth regulation will have in the market.

This housing regulation policy contradicts the intuition about a voiced general opinion that the land in Madrid should be deregulated to build more housing units and, therefore, decrease the overall prices. This suggestion might reduce the prices of the housing units, yet it could worsen the current situation
in Madrid in the long term. From the housing-demographic point of view, and according to this nonlinear model, 
more deregulation of land can push the system even farther out of equilibrium.

It is interesting to note that regulation policies for the lowest and the higher layers are closer to each other than the middle ones. Policies to regulate the poorest layer will also contribute positively in regulating the upper ones. The middle layers follow a different dynamics on its own mostly due to the fact that they receive and send family units from both sides of the distribution.

\section{Conclusion}
As pointed out by Brian Arthur \cite{arthur}:``...complexity economics, is not an adjunct to standard economic theory, but theory at a more general, out-of-equilibrium level.'' In this paper, we develop a novel multi-layered nonlinear dynamical framework for modeling the housing market dynamics. Using realistic data with the highest available precision, we show that the housing market for the city of Madrid is currently driving away from equilibrium. This model can be used as a testing tool to determine the global effects of policy changes by governments.  It is a tool to determine the effect of control parameters for all potential types of  dynamics even for  out-of-equilibrium conditions. Traditional econometric tools approximate the dynamics near set points, which are estimated (sometimes believed) to be the equilibrium points.  When the system gets out of equilibrium there is not much guidance about what to expect, except waiting till it gets near the equilibrium point again.
General tools to estimate effects of policy changes in the long run can be useful. Here we give an example that is able to provide 
a good prediction of the global level of occupancy in Madrid in 2001 (see Fig. \ref{fig5}A).

Our original goal was to find out whether the current house-market conditions in Madrid could lead to a critical condition 
such that, while slowly moving a parameter value as the temperature, the levels of occupancy
suddenly drop to  low levels. Fortunately, we did not find the existence of such criticality in the context of this
model that uses to the maximum possible extent realistic data. When the interest rate is increased, which reduces the 
temperature parameter in our model, the occupancy levels of the medium range housing are not  seriously
affected. The first layer suffers dramatic consequences, and the wealthier layers undergo serious readjustments.

Although this result appears to be good news (except for the poorest sector of the housing units), the fact is  that the city of Madrid is in a serious out-of-equilibrium condition that asymptotically drives the levels of occupancy to $0$. According to our model, interest rate corrections will not modify this condition. The pragmatical way to control this situation is to individually regulate the amount of new construction for each of the housing layers. Smooth changes in the construction rate can have a positive effect slowing down the out-of-equilibrium condition.

\section{Acknowledgments}
We want to thank Montserrat Mart\'{\i}nez and Cesar Pe\~nas for much of the data
mining. This work has been funded by Ministerio de Industria (Spain) PROFIT FIT 340000-2004-103.


\begin{thebibliography}{99}  

\bibitem[Arthur 1999]{arthur} 
Arthur, W.B. (1999).
Complexity and the economy.
Science, 284, 107-109

\bibitem[Bailey 1975]{epidemic} 
Bailey, N.T.J. (1975).
The mathematical theory of infectious diseases.
(London: Charles Griffin \& Co.)

\bibitem[BBVA 2003]{bbva} 
Banco Bilbao Vizcaya Argentaria, Servicio de Estudios (2003).
Situaci\'{o}n Inmobiliaria, Abril

\bibitem[Bin 2004]{pitt} 
Bin, O. (2004).
A prediction comparison of housing sale prices by parametric versus semi-parametric regressions.
Journal of Housing Economics, 13, 68-84

\bibitem[Crone and Mills 1991]{population2} 
Crone, T.M. \& Mills, L.O. (1991).
Forecasting trends in the housing stock using age-specific demographic projections. 
Journal of Housing Research, 2, 1-20

\bibitem[Garber 2000]{garber2000}
Garber P.M. (2000).
Famous first bubbles: the fundamentals of early manias.
(Cambridge, Massachusetts: The MIT Press) 
 
\bibitem[Green and Malpezzi 2003]{review} 
Green R.K. \& Malpezzi, S. (2003).
A primer on U.S. housing markets and policies.
(Washington DC: The Urban Institute Press) 

\bibitem[Huerta and Tsimring 2002]{huerta} 
Huerta, R. \& Tsimring, L.S. (2002).
Contact tracing and epidemics control in social networks.
Physical Review E, 66, 056115-4.

\bibitem[INE 1991]{censo91} 
Instituto Nacional de Estad\'{\i}stica (1991).
Censo de Poblaci\'{o}n y Viviendas

\bibitem[INE 2002]{epa}
Instituto Nacional de Estad\'{\i}stica (1987-2002).
Encuesta de Poblaci\'{o}n Activa

\bibitem[INE 2004]{ine} 
Instituto Nacional de Estad\'{\i}stica. \texttt{http://www.ine.es/}

\bibitem[Kindleberger 2000]{crashbook} 
Kindleberger, C.P. (2000). 
Manias, panics, and crashes: a history of financial crises. 
(New York: John Wiley \& Sons)

\bibitem[Linneman 1986]{regulation} 
Linneman, P. (1986).
An empirical test of the efficiency of the housing market.
Journal of Urban Economics, 20, 140-154

\bibitem[Malpezzi 1990]{review2} 
Malpezzi, S. (1990).
Urban housing and financial markets: some international comparisons.
Urban Studies, 27, 971-1022

\bibitem[Malpezzi 1999]{overprize} 
Malpezzi, S. (1999).
A simple error correction model of house prices.
Journal of Housing Economics, 8, 27-62

\bibitem[Mankiw and Weil 1989]{population1} 
Mankiw, G.N. \& Weil, D.N. (1989).
The baby boom, the baby bust, and the housing market. 
Regional Science and Urban Economics, 19, 235-58

\bibitem[OBJOVI 2004]{objovi} 
Observatorio Joven de la Vivienda en Espa\~{n}a\\ 
\texttt{http://www.cje.org/C14/C6/OBJOVI/default.aspx?lang=es-ES} 

\bibitem[Quigley 2001]{asian} 
Quigley, J.M. (2001).
Real state and the Asian crisis.
Journal of Housing Economics, 10, 129-161 

\bibitem[Sociedad de Tasaci\'{o}n S.A. 2004]{tasacion} 
Sociedad de Tasaci\'{o}n S.A.\\
\texttt{http://web.st-tasacion.es/html/index.php}

\bibitem[Terrones et al. 2004]{imf2} 
Terrones, M., Otrok, C. \& Carcenac, N. (2004).
The global house price boom. 
World Economic Outlook, chapter II, September 22, 2004, 71-89.
Retrieved December 20, 2004, from 
\texttt{http://www.imf.org/external/pubs/ft/weo/2004/02/}

\end{thebibliography}
\end{document}